\documentclass[runningheads]{llncs}

\usepackage{url}  
\usepackage{graphicx}  
\usepackage{subfigure}
\usepackage{algorithm}
\usepackage{algorithmic}

\usepackage{mathtools}
\usepackage{amsmath}
\usepackage{amsthm}
\usepackage{array}
\usepackage{amsfonts}
\usepackage{color}
\usepackage{amssymb}
\usepackage{diagbox}
\usepackage{bm}
\usepackage{multirow, makecell}

\begin{document}

\title{Enhancing Network Embedding with Auxiliary Information: An Explicit Matrix \\
Factorization Perspective}
\author{Junliang Guo
\and Linli Xu
\and Xunpeng Huang
\and Enhong Chen}
\institute{
Anhui Province Key Laboratory of Big Data Analysis and Application,\\
School of Computer Science and Technology,\\
University of Science and Technology of China \\
\email{guojunll@mail.ustc.edu.cn}, \email{linlixu@ustc.edu.cn} \\
\email{hxpsola@mail.ustc.edu.cn}, \email{cheneh@ustc.edu.cn}}

\titlerunning{Enhancing Network Embedding with Auxiliary Information}
\authorrunning{Junliang Guo \and Linli Xu \and Xunpeng Huang \and Enhong Chen}

\maketitle
\vspace{-0.1in}
\begin{abstract}
Recent advances in the field of network embedding have shown
the low-dimensional network representation is playing
a critical role in network analysis.
However, most of the existing principles of network embedding do
not incorporate auxiliary information such as content and labels of nodes flexibly.
In this paper, we take a matrix factorization perspective of network
embedding, and incorporate structure, content and label information
of the network simultaneously. For structure, we validate that the matrix we
construct
preserves high-order proximities of the
network. Label information can be further integrated into the matrix
via the process of random walk sampling to enhance the quality of embedding
in an unsupervised manner, i.e., without leveraging downstream classifiers.
In addition, we generalize the Skip-Gram Negative
Sampling model to integrate the content of the network in a
matrix factorization framework.
As a consequence, network embedding can be learned in a unified
framework integrating network structure and node content
as well as label information simultaneously.
We demonstrate the efficacy of the proposed model
with the tasks of semi-supervised node classification and link prediction on a
variety of real-world benchmark network datasets.

\vspace{-0.1in}
\end{abstract}

\section{Introduction}

The rapid growth of applications based on networks has
posed major challenges of effective processing of network data, among
which a critical task is network data representation. The primitive
representation of a network is usually very sparse and suffers from
overwhelming high dimensionality,
which limits its generalization in statistical learning. To
deal with this issue, network embedding aims to learn latent
representations of nodes on a network while preserving the structure
and the inherent properties of the network, which can be effectively
exploited by classical vector-based machine learning models for
tasks including node classification, link prediction, and community
detection, etc.~\cite{kipf2016semi,grover2016node2vec,liu2015community,cai2017comprehensive}.

Recently, inspired by the
advances of neural representation learning in language modeling,
which is based on the principle of learning the embedding vector of
a word  by predicting its
context~\cite{mikolov2013efficient,mikolov2013distributed},
a number of network embedding approaches have been proposed with the
paradigm of learning the embedding vector of a node by predicting
its neighborhood~\cite{perozzi2014,tang2015line,grover2016node2vec}. Specifically,
latent representations of network nodes are learned by treating short
random walk sequences as sentences to encode structural proximity in
a network. Existing results demonstrate the effectiveness of the
neural network embedding approaches in the tasks of node
classification, behavior prediction, etc.

However, existing network embedding methods, including DeepWalk~\cite{perozzi2014},
LINE~\cite{tang2015line} and node2vec~\cite{grover2016node2vec},
are typically based on
structural proximities only and do not incorporate other information
such as node content flexibly.
In this paper, we explore the
question whether network structure and auxiliary properties of the
network such as node content and label information can be integrated
in a unified framework of network embedding.
To achieve that,
we take a matrix
factorization perspective of network embedding with the
benefits of natural integration of structural embedding and content
embedding simultaneously,
where label information can be incorporated flexibly.

Specifically, motivated by the recent
work~\cite{li2015} that explains the word embedding model of
Skip-Gram Negative Sampling (SGNS) as a matrix factorization of the
words' co-occurrence matrix, we build a co-occurrence matrix of
structural proximities for a network based on a random walk sampling procedure.
The process of SGNS can then be formulated as minimizing a matrix
factorization loss, which can be naturally integrated with
representation learning of node content. In addition, label
information can be exploited in the process of building the
co-occurrence matrix to enhance the quality of network embedding,
which is achieved by decomposing the context of a node into the
structure context generated with random walks, as well as the label
context based on the given label information.

Our main contributions can be summarized as follows:
\begin{itemize}
    \setlength{\itemsep}{0pt}
    \setlength{\parsep}{0pt}
    \setlength{\parskip}{0pt}
\item We propose a
unified framework of
Auxiliary information Preserved Network Embedding with matrix factorization,
abbreviated as \emph{APNE},
which can
effectively learn the latent representations of nodes, and
provide a flexible integration of network structure, node content, as
well as label information
without leveraging downstream classifiers.
\item We verify that the structure matrix we generate is an approximation
of the high-order proximity of the network known as rooted PageRank.
\item We extensively evaluate our framework
on four benchmark datasets and two tasks including semi-supervised
classification and link prediction. Results show that the
representations learned by our proposed method are general and
powerful, producing significantly increased performance over the state
of the art on both tasks.
\end{itemize}

\section{Related Work}
\label{sec:related}

\subsection{Network Embedding}

Network embedding has been extensively studied in the
literature~\cite{cai2017comprehensive}.
Recently, motivated by the advances of neural representation
learning in language modeling, a number of embedding learning
methods have been proposed based on the Skip-Gram model. 
A representative model is DeepWalk~\cite{perozzi2014},
which exploits random walk to generate sequences of instances as the
training corpus, followed by utilizing the Skip-Gram model to obtain the
embedding vectors of nodes. 
Node2vec~\cite{grover2016node2vec} 
extend DeepWalk with sophisticated random walk schemes.
Similarly
in LINE~\cite{tang2015line} and GraRep~\cite{cao2015grarep},
network embedding is learned by directly optimizing the objective function
inspired from the Skip-Gram model.
To further incorporate auxiliary information into network embedding,
many efforts have been made. Among them, TADW~\cite{yang2015network}
formulates DeepWalk in a matrix
factorization framework, and jointly learns embeddings with the
structure information and preprocessed features of text information.
This work is further extended by
HSCA~\cite{zhang2016homophily}, DMF~\cite{zhang2016collective} and
MMDW~\cite{tu2016max} with various additional information.
In SPINE~\cite{guo2018spine}, structural identities are incorporated
to jointly preserve local proximity and global proximity of the network simultaneously in
the learned embeddings.
However, none of the above
models jointly consider structure, content and label information in a unified model,
which are the fundamental elements of a network~\cite{cai2017comprehensive}.
Recently, TriDNR~\cite{pan2016tri} and LANE~\cite{huang2017label}
tackle this problem both through implicit interactions between the three elements.
TriDNR leverages multiple skip-gram algorithms between node-word and word-label
pairs, while LANE first constructs three network affinity matrices from the three
elements respectively, followed by executing SVD on affinity matrices with
additional pairwise interactions.
Although empirically effective, these methods do not provide a
clear objective articulating how the three aspects
are integrated in the embeddings learned,
and is relatively inflexible to generalize to other scenarios.

In contrast to the above models, we propose a unified framework to
learn network embeddings from structure, content and label simultaneously.
The superiority of our framework is threefold: a) the structure matrix
we generate contains high-order proximities of the network, and the label
information is incorporated by explicitly manipulating the constructed matrix
rather than through implicit multi-hop interactions~\cite{pan2016tri,huang2017label}; b) instead of
leveraging label information through an explicitly learned classifier
(e.g., SVM~\cite{tu2016max}, linear classifier~\cite{zhang2016collective}
and neural networks~\cite{yang2016revisiting,kipf2016semi})
whose performance
is not only related to the quality of embeddings but also the specific classifiers being used, we exploit
the label information without leveraging any downstream classifiers,
which enables the flexibility of our model to generalize to different tasks;
c) while most of the above models only consider text descriptions of nodes, we use raw features contained
in datasets as the content information, which is more generalized to various
types of networks in real world such as social networks.

\subsection{Matrix Factorization and Word Embedding}
Matrix Factorization (MF) has been proven effective in various machine
learning tasks, such as dimensionality reduction,
representation
learning, recommendation systems, etc.
Recently, connections have been built between MF and word embedding
models. It is shown in~\cite{levy2014neural} that the Skip-Gram
with Negative Sampling (SGNS) model is an Implicit Matrix
Factorization (IMF) that factorizes a word-context matrix, where the
value of each entry is the pointwise mutual information (PMI)
between a word and context pair, indicating the strength of
association. It is further pointed out in~\cite{li2015} that the
SGNS objective can be reformulated in a representation learning view
with an Explicit Matrix Factorization (EMF) objective, where the
matrix being factorized here is the co-occurrence matrix among words
and contexts.

In this paper, we extend the matrix factorization perspective of
word embedding into the task of network embedding. More importantly,
we learn the network embedding by jointly factorizing the structure
matrix and the content matrix of the network, which can be further
improved by leveraging auxiliary label information. 
Different from most existing network embedding methods based on 
matrix factorization,
which employ either trivial objective functions (F-norm used in TADW) 
or traditional factorization algorithms (SVD used in GraRep) for optimization,
we design a novel objective function based on SGNS in our framework.
Furthermore,
the proposed method is general and
not confined to specific downstream tasks, such as link
prediction~\cite{grover2016node2vec} and node classification~\cite{perozzi2014},
and we do not leverage any classifiers either.

\section{Network Embedding with Matrix Factorization}
\label{sec:model}

\begin{figure*}[tb]
\centering
\centerline{\includegraphics[width=1.0\columnwidth]{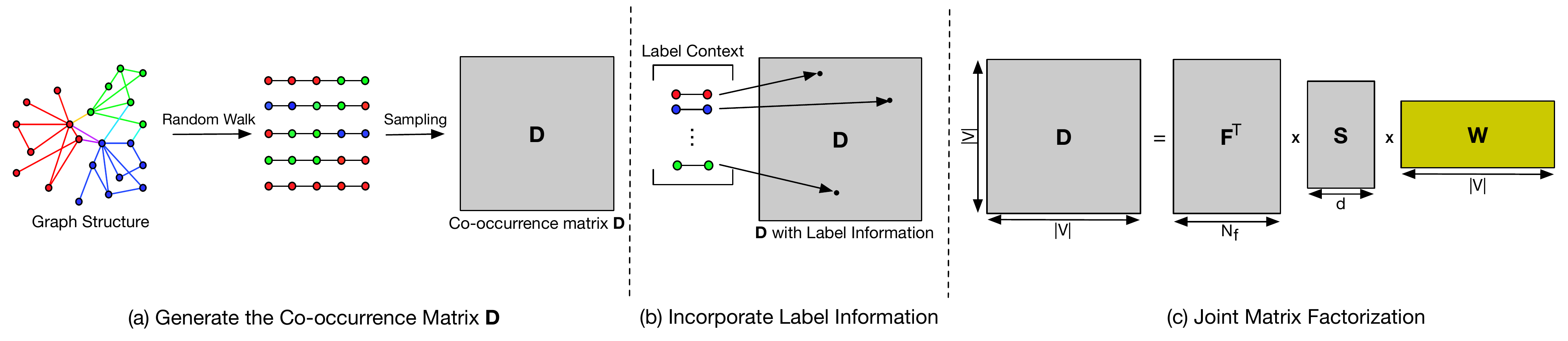}}
\caption{
The overall procedure of our framework (\emph{APNE}).
Different colors indicate
different labels of nodes.
}
\label{fig:process}
\vskip -0.1in
\end{figure*}

In this section, we propose a novel approach for network
embedding based on a unified matrix factorization framework, which consists of three procedures
as illustrated in Figure~\ref{fig:process}.
We
follow the paradigm of treating random walk sequences as sentences
to encode structural proximities in a network. However, unlike the
EMF objective for word embedding where the matrix to factorize is
clearly defined as the word-context co-occurrence matrix, for network
embedding, there is a gap between the random walk procedure and the
co-occurrence matrix. Therefore, we start with proposing a random
walk sampling process to build a co-occurrence matrix, followed by
theoretical justification of its property of preserving the
high-order structural proximity in the network, based on which we
present the framework of network embedding with
matrix factorization.

\subsection{High-Order Proximity Preserving Matrix}
\label{sec:theorem}

Given an undirected network
$G=\{V,E\}$ which includes a set of nodes $V$ connected by a set of
edges $E$, the corresponding adjacency matrix is $\bm{A}$, where
$A_{i,j} = w_{i,j}$ indicates an edge with weight $w_{i,j}$ between the $i$-th node $v_{i}$
and the $j$-th node $v_{j}$. And we denote the transition matrix of $G$ as $\bm{P}$,
where $P_{i,j} = \frac{w_{i,j}}{\sum_{k=1}^{|V|} w_{i,k}}$.
Next, a list of node sequences $C$ can
be generated with random walks on the network.

Given $C$, we can generate the co-occurrence
matrix $\bm{D}$ of $G$ with the $n$-gram algorithm.
The procedure is summarized in Algorithm~\ref{alg:A0}.
In short, for a given node in a node sequence, we increase
the co-occurrence count of two nodes if and only if they are
in a window of size $l$.
\begin{algorithm}[tb]
    \caption{Sampling the general co-occurrence matrix}
    \label{alg:A0}
\begin{algorithmic}[1]
    \REQUIRE The transition matrix $\bm{P}$, window size $l$
    \ENSURE Co-occurrence matrix $\bm{D}$
    \STATE Sample random walks $C$ based on $\bm{P}$
    \FOR {every node sequence in $C$}
    \STATE Uniformly sample $(i, j)$ with $|i - j| < l$
    \STATE $D_{v_{i}, v_{j}} = D_{v_{i}, v_{j}} + 1$
    \ENDFOR
\end{algorithmic}
\end{algorithm}

Next we show that the co-occurrence matrix generated by
Algorithm~\ref{alg:A0} preserves the high-order structural
proximity in the network with the following theorem:

\begin{theorem}
\label{theo:co-occ} Define the high-order proximity $\bm{S}$ of the
network $G$ as
\begin{center} \small
$\bm{S^{l}} = \sum_{k=1}^{l} \bm{P}^{k} $
\end{center}
where $l$ denotes the order of the proximity as well as the window size in Algorithm~\ref{alg:A0}.
Then, under the
condition that the random walk procedure is repeated enough times
and the generated list of node sequences $C$ covers all paths in the
network $G$, we can derive that according to~\cite{yang2015network}:
\begin{equation} \small
\label{equ:normal_D}
l \cdot \bm{D}^{nor} = \bm{S^{l}}
\end{equation}
where $l$ is the window size in Algorithm~\ref{alg:A0}, and the matrix
$\bm{D}^{nor}$ denotes the expectation of row normalized co-occurrence matrix $\bm{D}$, i.e.,
$\bm{D}^{nor}_{i,j} = \mathbb{E} [ \frac{\bm{D}_{i,j}}{\sum_{k=1}^{|V|} D_{i,k}}]$.
\end{theorem}

Note that
the $(i,j)$-th entry of the left side of
Equation~(\ref{equ:normal_D}) can be written as
$\mathbb{E} [\frac{\bm{D}_{i,j}}{\sum_{k=1}^{|V|} D_{i,k} / l}]$, which is the
expected number of times that $v_{j}$ appears in the left or right
$l$-neighborhood of $v_{i}$.

To investigate into the structural information of the network encoded in the co-occurrence matrix $\bm{D}$,
we first consider a well-known high-order proximity of a network named rooted PageRank (RPR)~\cite{song2009scalable},
defined as $\bm{S}^{\textrm{RPR}} = (1-\beta_{\textrm{RPR}})(\bm{I}-\beta_{\textrm{RPR}}\bm{P})^{-1}$, where $\beta_{\textrm{RPR}}\in (0,1)$ is
the probability of randomly walking to a neighbor rather than jumping back. The $(i,j)$-th entry of
$\bm{S}^{\textrm{RPR}}$ is the probability that a random walk from node $v_{i}$ will stop at $v_{j}$
in the steady state, which can be used as an indicator of the node-to-node proximity.
$\bm{S}^{\textrm{RPR}}$ can be further rewritten as:
\begin{small}
\begin{align}
\label{def:RPR}
\bm{S}^{\textrm{RPR}} = &(1-\beta_{\textrm{RPR}})(\bm{I}-\beta_{\textrm{RPR}}\bm{P})^{-1} \\
= &(1-\beta_{\textrm{RPR}})\sum_{k=0}^{\infty}\beta_{\textrm{RPR}}^{k}\bm{P}^{k}
\end{align}
\end{small}

We next show that for an undirected network, where $\bm{P}$ is symmetric,
the row normalized co-occurrence matrix $\bm{D}^{nor}$
is an approximation of the rooted PageRank matrix $\bm{S}^{\textrm{RPR}}$.
\begin{theorem}
\label{theo:appro}
When $l$ is sufficiently large, for $\bm{D}^{nor}$ defined as $\bm{D}^{nor}=\frac{1}{l}\sum_{k=1}^{l}\bm{P}^{k}$,
and $K = \lfloor -\frac{\log l(1-\beta_{\textrm{RPR}})}{\log \beta_{\textrm{RPR}}}\rfloor$,
the $\ell$-2 norm of the difference between $\bm{D}^{nor}$ and $\bm{S}^{\textrm{RPR}}$
can be bounded by $K$:
\begin{equation} \small
\label{equ:bounded}
\left \lVert \bm{S}^{\textrm{RPR}} - \bm{D}^{nor} \right \rVert _{2} \leq 2-2\beta_{\textrm{RPR}}^{K+1}
\end{equation}
\end{theorem}

\noindent \textbf{Proof of Theorem~\ref{theo:appro}.}~~~
Here we omit the superscript
of $\bm{S}^{\textrm{RPR}}$ and the subscript of $\beta_{\textrm{RPR}}$ in the proof for simplicity.
Substituting (\ref{def:RPR}) and reformulating the left side of (\ref{equ:bounded})
we have:
\begin{small}
\begin{align*}
&\left \lVert \bm{S} - \bm{D}^{nor} \right \rVert _{2} =
\left \lVert (1-\beta)\sum_{k=0}^{\infty}\beta^{k}\bm{P}^{k} - \frac{1}{l}\sum_{k=1}^{l}\bm{P}^{k} \right \rVert_{2}\\
&=\left \lVert (1-\beta)\sum_{k=0}^{l}\beta^{k}\bm{P}^{k} - \frac{1}{l}\sum_{k=0}^{l}\bm{P}^{k} + \frac{1}{l}
+(1-\beta)\sum_{k=l+1}^{\infty}\beta^{k}\bm{P}^{k}\right \rVert _{2} \\
&\leq \left \lVert \sum_{k=0}^{l}\bm{P}^{k}\left [ (1-\beta)\beta^{k} - \frac{1}{l} \right ]\right \rVert _{2}
+ (1-\beta)\left \lVert \sum_{k=l+1}^{\infty}\beta^{k}\bm{P}^{k}\right \rVert _{2} + \frac{1}{l}\\
&\leq \sum_{k=0}^{l} \lambda_{\max}^{k} \left |(1-\beta)\beta^{k} - \frac{1}{l} \right |
+(1-\beta)\sum_{k=l+1}^{\infty}\beta^{k}\lambda_{\max}^{k}+ \frac{1}{l}
\end{align*}
\end{small}
where $\lambda_{\max}$ is the largest singular value of matrix $\bm{P}$, which is also
the eigenvalue of $\bm{P}$ for the reason that $\bm{P}$ is symmetric and non-negative.
Note that $\bm{P}$ is the transition matrix, which is also known as the Markov matrix.
And it can be easily proven that the largest eigenvalue of a Markov matrix is always $1$,
i.e., $\lambda_{\max} = 1$.
We eliminate the absolute value sign by splitting the summation at
$K= \lfloor -\frac{\log l(1-\beta)}{\log \beta}\rfloor$, then we have:
\begin{small}
\begin{align*}
\left \lVert \bm{S} - \bm{D}^{nor} \right \rVert _{2} &\leq \sum_{k=0}^{K} \left [(1-\beta)\beta^{k} - \frac{1}{l} \right]
+ \sum_{k=K+1}^{l} \left [\frac{1}{l} - (1-\beta)\beta^{k}\right]\\
&+(1-\beta)\sum_{k=l+1}^{\infty}\beta^{k} + \frac{1}{l}\\
&=1-\beta^{K+1} - \beta^{K+1}(1-\beta^{l-K}) + \frac{l-2K}{l}+\beta^{l+1} \\
&= 1-2\beta^{K+1} + 2\beta^{l+1} + \frac{l-2K}{l}
\end{align*}
\end{small}
Note that when $l$ is sufficiently large, according to the definition of $K$, we have $K \ll l$.
Given $\beta \in (0,1)$, we can derive:
\begin{equation*} \small
\left \lVert \bm{S} - \bm{D}^{nor} \right \rVert _{2} \leq 2-2\beta^{K+1}.
\end{equation*}
$\hfill \qed $

With Theorem~\ref{theo:appro}, we can conclude that the normalized
co-occurrence matrix $\bm{D}^{nor}$
we construct is an approximation of the rooted PageRank matrix $S^{\textrm{RPR}}$ with a bounded $\ell$-2 norm.

Note that in TADW~\cite{yang2015network}
and its follow-up works~\cite{zhang2016homophily,zhang2016collective}
which also apply matrix
factorization to learn network embeddings, the matrix constructed to
represent the structure of a network is $\frac{\bm{P}+\bm{P}^2}{2}$,
which is a special case of $\bm{D}^{nor}$ when $l=2$. As comparison,
we construct a general matrix while preserving high-order
proximities of the network with theoretical justification.

\subsection{Incorporating Label Context}
\label{sec:label}

\begin{algorithm}[tb]
    \caption{Sampling general co-occurrence matrix with structure and label context}
    \label{alg:A}
\begin{algorithmic}[1]
    \REQUIRE The transition matrix $\bm{P}$, labeled nodes $L$, parameters $m$, $l$, $d$
    \ENSURE Co-occurrence matrix $\bm{D}$
    \STATE Sample random walks ${C}$ of length $d$ 
    based on $\bm{P}$
    \FOR {every node sequence in $C$}
    \STATE Uniformly sample $(i, j)$ with $|i - j| < l$
    \STATE $D_{v_{i}, v_{j}} = D_{v_{i}, v_{j}} + 1$
    \ENDFOR
    \FOR{$k=1$ to $m$}
    \STATE Uniformly sample a node $v_i$ in $L$
    \STATE Uniformly sample a node $v_j$ with the same label as node $v_i$
    \STATE $D_{v_{i}, v_{j}} = D_{v_{i}, v_{j}} + 1$
    \ENDFOR
\end{algorithmic}
\end{algorithm}

Apparently, the co-occurrence value between node $v_i$ and context $v_c$
indicates the similarity between them. A larger value of
co-occurrence indicates closer proximity in the network, hence higher
probability of belonging to the same class. 
This intuition coincides with the label information
of nodes. Therefore, with the benefit of integer values in $\bm{D}$,
label information can be explicitly incorporated in the procedure of sampling
$\bm{D}$ to enhance the proximity between nodes, which can additionally alleviate
the problem of isolated nodes without co-occurrence in structure,
i.e., we consider isolated nodes through label context instead of structure context.

Specifically, we randomly sample one node
among labeled instances, followed by uniformly choosing another node
with the same label and update the corresponding co-occurrence count
in
$\bm{D}$. As a consequence, the co-occurrence matrix $\bm{D}$
captures both structure co-occurrence and label co-occurrence of
instances. The complete procedure is summarized in Algorithm
\ref{alg:A}, where  $m$ is a parameter controlling the ratio
between the structure and label context.

In this way, while preserving high-order proximities of the network,
we can incorporate supervision into the model flexibly without
leveraging any downstream classifiers, which is another important
advantage of our method.  
By contrast, most existing methods are
either purely 
unsupervised~\cite{yang2015network}
or leveraging label information through downstream
classifiers~\cite{tu2016max,zhang2016collective}.

\subsection{Joint Matrix Factorization}
The method proposed above generates the co-occurrence matrix from a
network and bridges the gap between word embedding and network
embedding, allowing us to apply the matrix factorization paradigm to
network embedding. 
With the
flexibility of the matrix factorization principle, we propose a
joint matrix factorization model that can learn network embeddings
exploiting not only the topological structure and label information but also the content
information of the network simultaneously.

Given the co-occurrence matrix $\bm{D} \in \mathbb{R}^{|V| \times
|V|}$ and the content matrix $\bm{F} \in \mathbb{R}^{N_{f} \times |V|}$,
where $|V|$ and $N_{f}$ represent the number of nodes in the network
and the dimensionality of node features respectively. 
Let $d$ be the
dimensionality of embedding. The 
objective here is to learn the
embedding of a network $G$, denoted as the matrix $\bm{W} \in
\mathbb{R}^{d \times |V|}$,
by minimizing the loss of
factorizing the matrices $\bm{D}$ and $\bm{F}$
jointly as:
\begin{equation} \small
\label{def:joint-mf}
\min_{\bm{W},\bm{S}} MF(\bm{D}, \bm{F}^{T}\bm{S}\bm{W})
\end{equation}
where $MF(\cdot, \cdot)$ is
the reconstruction loss of matrix factorization
which will be introduced later,
and $\bm{S} \in \mathbb{R}^{N_{f}
\times d}$ can be regarded as the \emph{feature embedding} matrix,
thus $\bm{F}^{T}\bm{S}$ is the \emph{feature embedding} dictionary of
nodes.

By solving the joint matrix factorization problem in
(\ref{def:joint-mf}), the structure information in $\bm{D}$ and the
content information in $\bm{F}$ are integrated to learn the network
embeddings $\bm{W}$. This is inspired by Inductive Matrix
Completion~\cite{natarajan2014inductive}, a method originally
proposed to complete a gene-disease matrix with gene and disease
features. However, we take a completely different loss function here
in light of the word embedding model of SGNS with a matrix
factorization perspective~\cite{li2015}.

We first rewrite (\ref{def:joint-mf}) in a representation learning
view as:
\begin{equation} \small
\label{def:joint-rl}
\min_{\bm{W},\bm{S}} \sum_{i} MF(\bm{d_{i}}, \bm{F}^{T}\bm{S}\bm{w_{i}})
\end{equation}
where $MF(\cdot,\cdot)$ is the representation loss functions
evaluating the discrepancy between the $i^{th}$ column of $\bm{D}$
and $\bm{F}^{T}\bm{S}\bm{W}$.
$\bm{F}^{T}\bm{S}$ is the feature
embedding dictionary, and the embedding vector of the $i^{th}$
node, $\bm{w_{i}}\in \mathbb{R}^{d}$, can be learned by
minimizing the loss of representing its structure context vector
$\bm{d_{i}}$ via the feature embedding $\bm{F}^{T}\bm{S}$.

We then proceed to the objective of factorizing the co-occurrence matrix $\bm{D}$ and
the content matrix $\bm{F}$ jointly,
denoted as
$MF(\bm{d_{i}}, \bm{F}^{T}\bm{S}\bm{w_{i}})$. We follow the paradigm
of explicit matrix factorization of the SGNS model and derive the following theorem according to~\cite{li2015}:

\begin{theorem}
\label{theo:emf}
For a node $i$ in the network, we denote $Q_{i,c}$ as a pre-defined upper bound for the possible
co-occurrence count between node $i$ and context $c$.
With the equivalence of Skip-Gram Negative Sampling (SGNS) and Explicit Matrix Factorization (EMF)
\cite{li2015},
the representation loss $MF(\cdot,\cdot)$ can be defined as the negative log
probability of observing the structure vector $\bm{d_{i}}$ given $i$ and $\bm{F}^{T}\bm{S}$
when $Q_{i,c}$ is set to $k \frac{\#(i)\#(c)}{|D|} + \#(i,c)$. To be more concrete,
\begin{center} \small
$MF(\bm{d_{i}}, \bm{F}^{T}\bm{S}\bm{w_{i}})= - \sum_{c \in |V|} \log P(d_{i,c}|\bm{f}^{T}_c\bm{S}\bm{w_i})$
\end{center}
where $\bm{f}_{c}\in \mathbb{R}^{N_{f}}$ is the $c$-th column of the content matrix $\bm{F}$, i.e., the feature vector
of node $c$,
$\#(i,c)$ is the co-occurrence count between node $i$ and $c$, $\#(i)=\sum_{c \in |V|}\#(i,c)$,
$\#(c) = \sum_{i \in |V|} \#(i,c)$, $|D| = \sum_{i,c \in |V|} \# (i,c)$ and $k$ is the
negative sampling ratio.
\end{theorem}

Based on Theorem~\ref{theo:emf}, we can derive:
\begin{equation*} \small
\label{def:mf-d}
\begin{aligned}
MF(\bm{D}, \bm{F}^{T}\bm{S}\bm{W}) &\triangleq \sum_{i=1}^{|V|} MF(\bm{d_{i}}, \bm{F}^{T}\bm{S}\bm{w_{i}}) \\
&= - \sum_{i=1}^{|V|} \sum_{c=1}^{|V|} \log P(d_{i,c}|\bm{f}^{T}_c\bm{S}\bm{w_i})
\end{aligned}
\end{equation*}

Finally, we can formulate the objective of the joint matrix
factorization framework with parameters $\bm{W}$ and $\bm{S}$ as:
\begin{equation} \small
\label{equ:total-loss}
\begin{aligned}
L(\bm{W},\bm{S}) &= MF(\bm{D}, \bm{F}^{T}\bm{S}\bm{W}) \\
&=- \sum_{i=1}^{|V|} \sum_{c=1}^{|V|} \log P(d_{i,c}|\bm{f}^{T}_c\bm{S}\bm{w_i})
\end{aligned}
\end{equation}

\subsection{Optimization}

To
minimize the loss function in (\ref{equ:total-loss}) 
which integrates structure, label and content simultaneously, we 
utilize
a novel
optimization algorithm leveraging the alternating minimization
scheme (ALM), which is a widely adopted method in the matrix
factorization literature.

First we derive the gradients of (\ref{equ:total-loss}) as:
\begin{equation*} \small
\begin{aligned}
\frac{\partial L(\bm{W},\bm{S})}{\partial \bm{S}} &= \frac{\partial MF(\bm{D}, \bm{F}^{T}\bm{S}\bm{W})}{\partial \bm{S}}  \\
&= \sum_{i \in |V|} - \bm{d_{i}}\bm{w_{i}}^{T} + \mathbb{E}_{\bm{d_{i}}^{'}|\bm{F}^{T}\bm{S}\bm{w_{i}}}[\bm{d_{i}}^{'}]\bm{w_{i}}^{T} \\
&= \bm{F}(\mathbb{E}_{\bm{D}^{'}|\bm{F}^{T}\bm{S}\bm{W}}\bm{D}^{'} - \bm{D})\bm{W}^{T}\\
& \triangleq \mathrm{grad}_{\bm{S}}\\
\frac{\partial L(\bm{W},\bm{S})}{\partial \bm{W}} &= \bm{S}^{T}\bm{F}(\mathbb{E}_{\bm{D}^{'}|\bm{F}^{T}\bm{S}\bm{W}}\bm{D}^{'} - \bm{D}) \\
&\triangleq \mathrm{grad}_{\bm{W}}
\end{aligned}
\end{equation*}

We denote $\mathrm{grad}_{\bm{W}}$ and $\mathrm{grad}_{\bm{S}}$ as
the gradients of $\bm{W}$ and
$\bm{S}$ in the loss
function~(\ref{equ:total-loss}) respectively. Note that the
expectation $\mathbb{E}_{\bm{d_{i}}^{'}|\bm{F}^{T}\bm{S}\bm{w_i}}$
can be computed in a closed form \cite{li2015} as:
\begin{equation} \small
\label{equ:closed}
\mathbb{E}_{d_{i,c}^{'}|\bm{f}^{T}_c\bm{S}\bm{w_i}}[d_{i,c}^{'}]=Q_{i,c}\sigma(\bm{f}^{T}_c\bm{S}\bm{w_i})
\end{equation}
where $\sigma (x)=\frac{1}{1+e^{-x}}$ is the sigmoid function.

The algorithm of Alternating Minimization (ALM) is summarized in
Algorithm~\ref{alg:op}. The algorithm can be divided into solving
two convex subproblems (starting from line 3 and line 6
respectively), which guarantees that the optimal solution of each
subproblem can be reached with sublinear convergence rate with a
properly chosen step-size~\cite{nesterov2013introductory}. 
One
can easily show that the objective~(\ref{equ:total-loss}) descents
monotonically.
As a consequence,
Algorithm~\ref{alg:op} will converge due to the lower bounded
objective function~(\ref{equ:total-loss}).

The time complexity of one iteration in Algorithm~\ref{alg:op} is $O((nnz(\bm{F})d|V|)^2)$,
where $nnz(\bm{F})$ is the number of non-zero elements in $\bm{F}$. 
For datasets with sparse node content, e.g., Cora, Citeseer, Facebook, etc.,
we implement $\bm{f}^{T}_c\bm{S}$ in Equation~(\ref{equ:closed}) efficiently as a product of a sparse matrix with a dense matrix,
which reduces the complexity from $O(|V|N_{f}d)$ to $O(nnz(\bm{F})d)$.

\begin{algorithm}[tb]
\caption{ALM algorithm for generalized explicit matrix factorization}
\label{alg:op}
\begin{algorithmic}[1]
\REQUIRE Co-occurrence matrix $\bm{D}$, content matrix $\bm{F}$, ALM step-size $\mu$ and
maximum number of outer iterations $I$
\ENSURE Node embedding matrix $\bm{W}$, feature embedding dictionary $\bm{S}$
\STATE Initialize $\bm{W}$ and $\bm{S}$ randomly
\FOR{$i=1$ to $I$}
\REPEAT
\STATE $\bm{W}=\bm{W} - \mu \cdot \mathrm{grad}_{\bm{W}}$
\UNTIL{Convergence}
\REPEAT
\STATE $\bm{S} = \bm{S}- \mu \cdot \mathrm{grad}_{\bm{S}}$
\UNTIL{Convergence}
\ENDFOR
\end{algorithmic}
\end{algorithm}

\section{Experiments}
\label{experiments}
The proposed framework is independent of specific downstream tasks, 
therefore in experiments, we test the model with different tasks 
including link prediction and node classification. Below we first 
introduce the datasets we use and the baseline methods that we compare to.

\begin{table}[tb]
\centering
\caption{Dataset statistics}
\label{tab:data}
\vskip -0.1in
\begin{tabular}{p{2cm} <{\centering}p{2cm} <{\centering}p{2cm} <{\centering}p{2cm} <{\centering}p{2cm} <{\centering}}
\hline
Dataset&\# Classes&\# Nodes&\# Edges&\# Feature\\
\hline
\hline
Citeseer & $6$ & $3327$ & $4732$ & $3703$ \\
Cora & $7$ & $2708$ & $5429$ & $1433$ \\
Pubmed & $3$ & $19717$ & $44338$ & $500$ \\
Facebook & - & $4309$ & $88234$ & $1283$ \\
\hline
\end{tabular}
\vskip -0.1in
\end{table}

\noindent \textbf{Datasets.}~
We test our models on four benchmark datasets. The statistics of datasets
are summarized in Table \ref{tab:data}. 
For the node classification task, we employ datasets of
Citation Networks~\cite{sen2008collective}, where nodes represent papers while edges represent citations.
And each paper is described by a one-hot vector or a TFIDF word vector.
For the link prediction task, we additionally include a social network dataset Facebook~\cite{snapnets}.
This dataset consists of $10$ ego-networks from the online
social network Facebook, where nodes and edges represent users and their relations respectively.
Each user is described by users' properties, which is represented by a one-hot vector.

\noindent \textbf{Baselines.}~
For both tasks, we compare our method with network embedding algorithms including 
DeepWalk~\cite{perozzi2014}, node2vec~\cite{grover2016node2vec},
TADW~\cite{yang2015network} and HSCA~\cite{zhang2016homophily}. For the node classification task, 
we further include DMF~\cite{zhang2016collective}, LANE~\cite{huang2017label} and
two neural network based methods,
Planetoid~\cite{yang2016revisiting} and GCN~\cite{kipf2016semi}.
To measure the performance of link prediction, we also evaluate
our method against some popular heuristic scores defined 
in node2vec~\cite{grover2016node2vec}.
Note that we do not consider TriDNR~\cite{pan2016tri} as a baseline
for the reason that they use text description as node content in citation networks, while
in social networks such as Facebook, there is no natural text description for each user,
which prevents TriDNR from generalizing to various types of networks.
In addition, as MMDW~\cite{tu2016max} and DMF~\cite{zhang2016collective} are both
semi-supervised variants of TADW with similar performance in our setting, we only compare our model
with DMF for brevity.

\noindent \textbf{Experimental Setup.}~
For our model, the hyper-parameters are tuned on the Citeseer dataset and kept on the others.
The dimensionality of embedding
is set to $200$ for the proposed methods.
In terms of the optimization parameters, the number of iterations is set to $200$,
the step-size in
Algorithm~\ref{alg:op} is set to $1e-7$.
The parameters in  Algorithm \ref{alg:A} are set in consistency with DeepWalk,
i.e., walk length $d = 40$ with window size $l = 5$. We use 
\emph{APNE}
to denote our unsupervised model of network embedding where  the co-occurrence matrix
is generated by Algorithm~\ref{alg:A0}, and
\emph{APNE+label}
denotes the semi-supervised model which uses
Algorithm~\ref{alg:A} to incorporate label context into the co-occurrence matrix.
Unless otherwise specified,
in all the experiments, we use one-vs-rest logistic regression as the classifier
for the embedding based methods\footnote{Code available at \url{https://github.com/lemmonation/APNE}}.

\subsection{Semi-supervised Node Classification}

\begin{table}[tb]
\centering
\caption{Accuracy of semi-supervised node classification (in percentage).
Upper and lower rows correspond to unsupervised and semi-supervised embedding methods respectively.
}
\label{tab:results}
\vskip -0.1in
\begin{tabular}{p{2.2cm} <{\centering}||p{1.5cm} <{\centering}p{1.5cm} <{\centering}p{1.5cm} <{\centering}}
\hline
Method & Citeseer & Cora & Pubmed \\
\hline
\hline
DeepWalk & $41.5$ & $67.3$ & $66.4$ \\
node2vec & $47.2$ & $69.8$ & $70.3$ \\
TADW & $54.0$ & $72.0$ & $41.7$ \\
HSCA & $47.7$ & $65.4$ & $41.7$ \\
\emph{APNE} & $\bm{72.6}$ & $\bm{79.3}$ & $\bm{81.5}$ \\
\hline
DMF & $65.5$ & $58.5$ & $59.3$ \\
LANE & $60.3$ & $65.2$ & - \\
Planetoid & $67.3$ & $73.4$ & $76.7$ \\
GCN & $70.3$ & $\bm{81.5}$ & $79.0$ \\
\emph{APNE+label} & $\bm{72.8}$ & $79.6$ & $\bm{82.1}$ \\
\hline
\end{tabular}
\vskip -0.25in
\end{table}

We first consider the semi-supervised node classification task on three citation network datasets.
To facilitate the comparison between our model and the baselines, we use the same partition
scheme of training set and test set as in \cite{yang2016revisiting}. To be concrete,
we randomly sample $20$ instances from each
class as training data, and $1000$ instances from all samples in the rest of the dataset as test data.

The experimental results are reported in Table \ref{tab:results}.
In the comparison of unsupervised models,
the proposed \emph{APNE} method learns embeddings from the network structure and node
content jointly in a unified matrix factorization framework. 
As a consequence, \emph{APNE} outperforms notably on all datasets
with improvement from $10.1\%$ to $34.4\%$.
Compared with TADW and HSCA, which both incorporate network topology and text features of nodes
in a matrix factorization
model simultaneously, our method is superior in the following:  
a) the matrix we construct and factorize represents the network topology better
as proven in Section~\ref{sec:theorem};
b) the loss function we derive from SGNS is tailored for representation learning.

Meanwhile, in the comparison of semi-supervised methods, the proposed \emph{APNE}
model outperforms embedding based baselines significantly, 
illustrating
the promotion brought by explicitly manipulating the constructed matrix
rather than implicitly executing multi-hop interactions.
In addition, LANE suffers from extensive complexity both in time and space,
which prevents it from being generalized to larger networks such as Pubmed.
Although being slightly inferior to GCN
on the Cora dataset, considering that \emph{APNE} is a feature learning method
independent of downstream tasks and classifiers, the competitive results against
the state-of-the-art CNN based method GCN justify that the node representations learned
by \emph{APNE} preserve the network information well.

In general, the proposed matrix factorization
framework outperforms embedding based baselines and
performs competitive with the state-of-the-art CNN based model,
demonstrating the quality of embeddings learned 
by our methods to represent the network from the aspects of content and structure. 
Between the two variants of our proposed framework, \emph{APNE} and \emph{APNE+label},
the latter performs consistently better on all datasets,
indicating the benefits of incorporating label context.

\begin{figure*}[tb]
\centering
\subfigure[\emph{APNE}]{\label{fig:cora_mf}\includegraphics[width=0.3\textwidth]{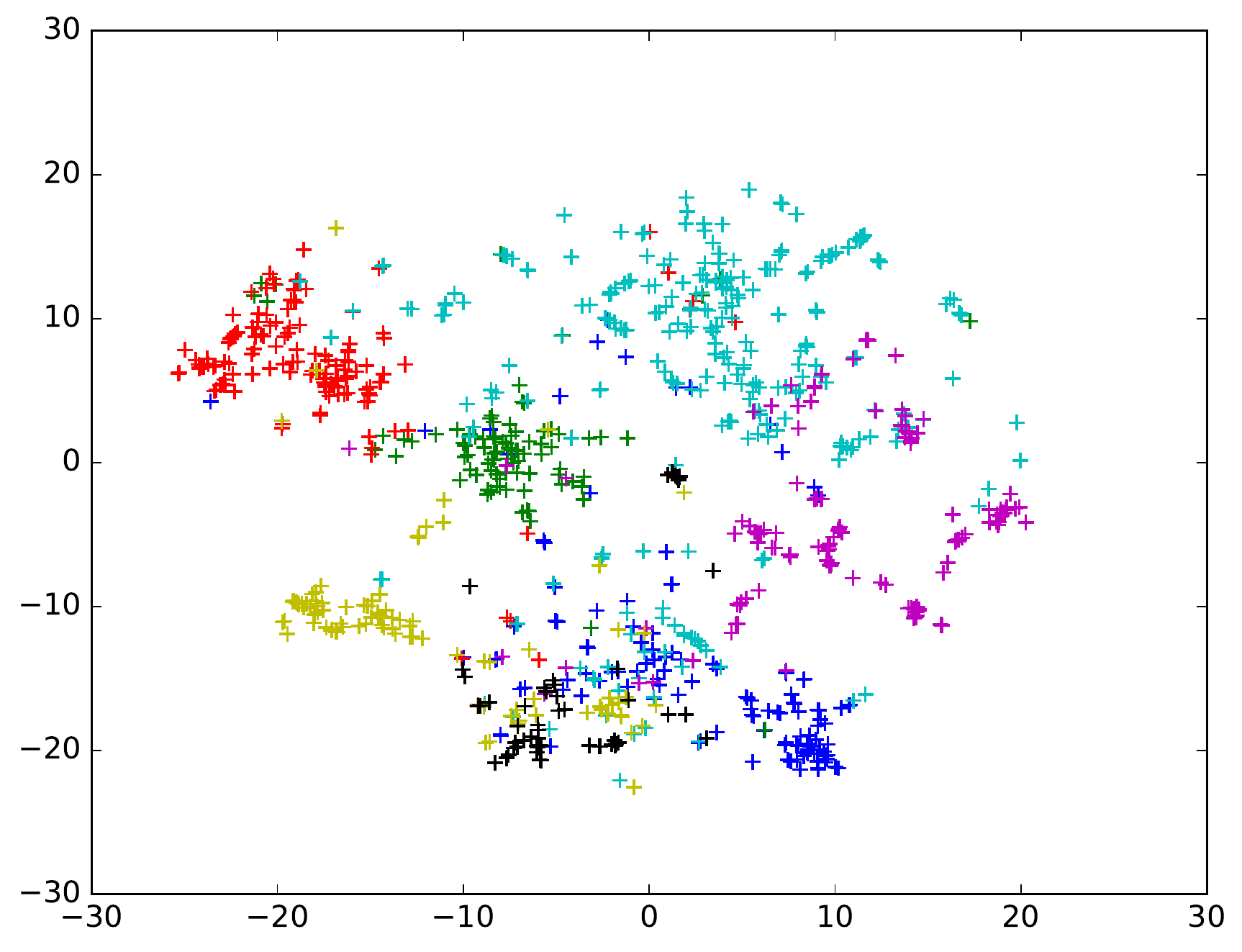}}
\subfigure[node2vec]{\label{fig:cora_node2vec}\includegraphics[width=0.3\textwidth]{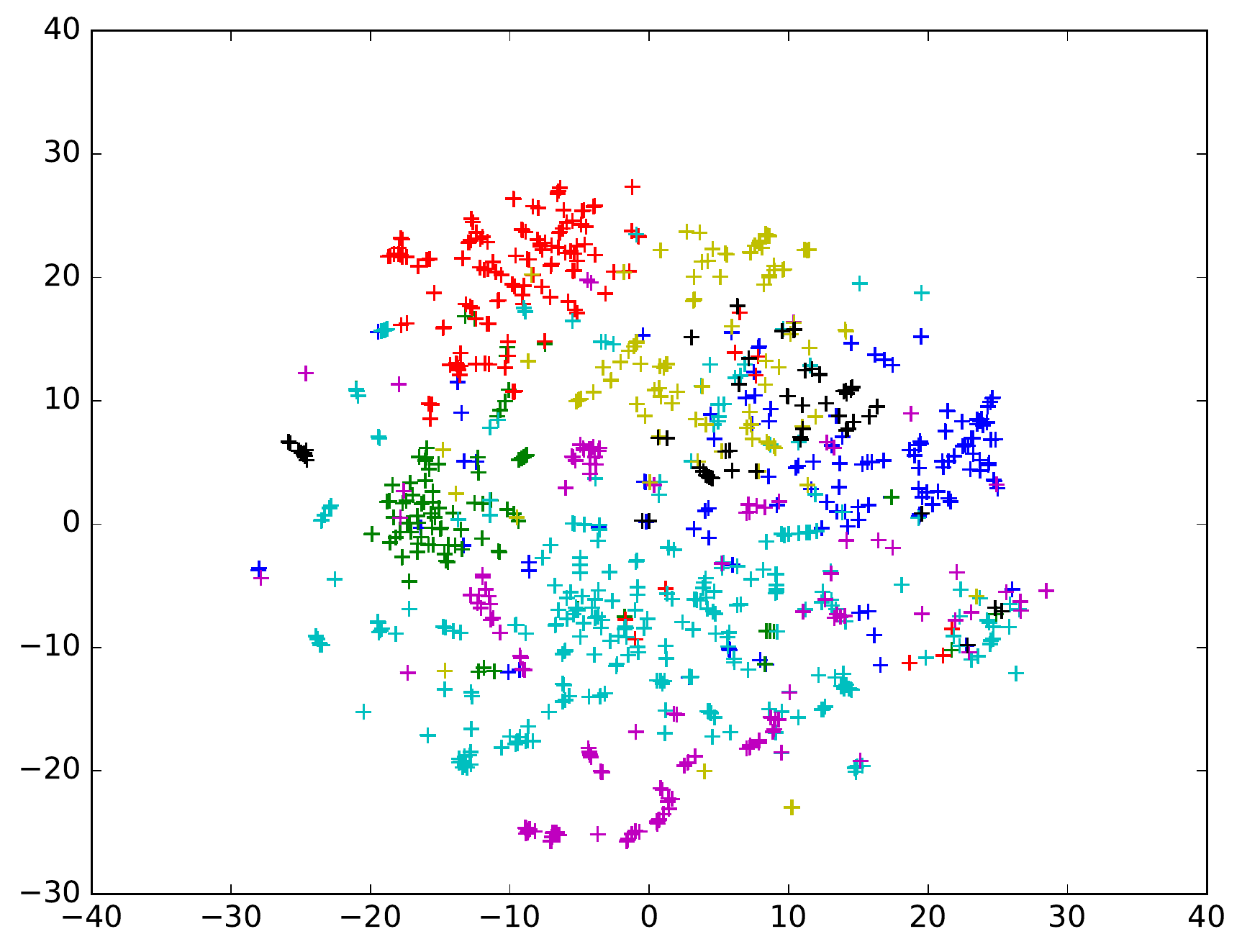}}
\subfigure[TADW]{\label{fig:cora_TADW}\includegraphics[width=0.3\textwidth]{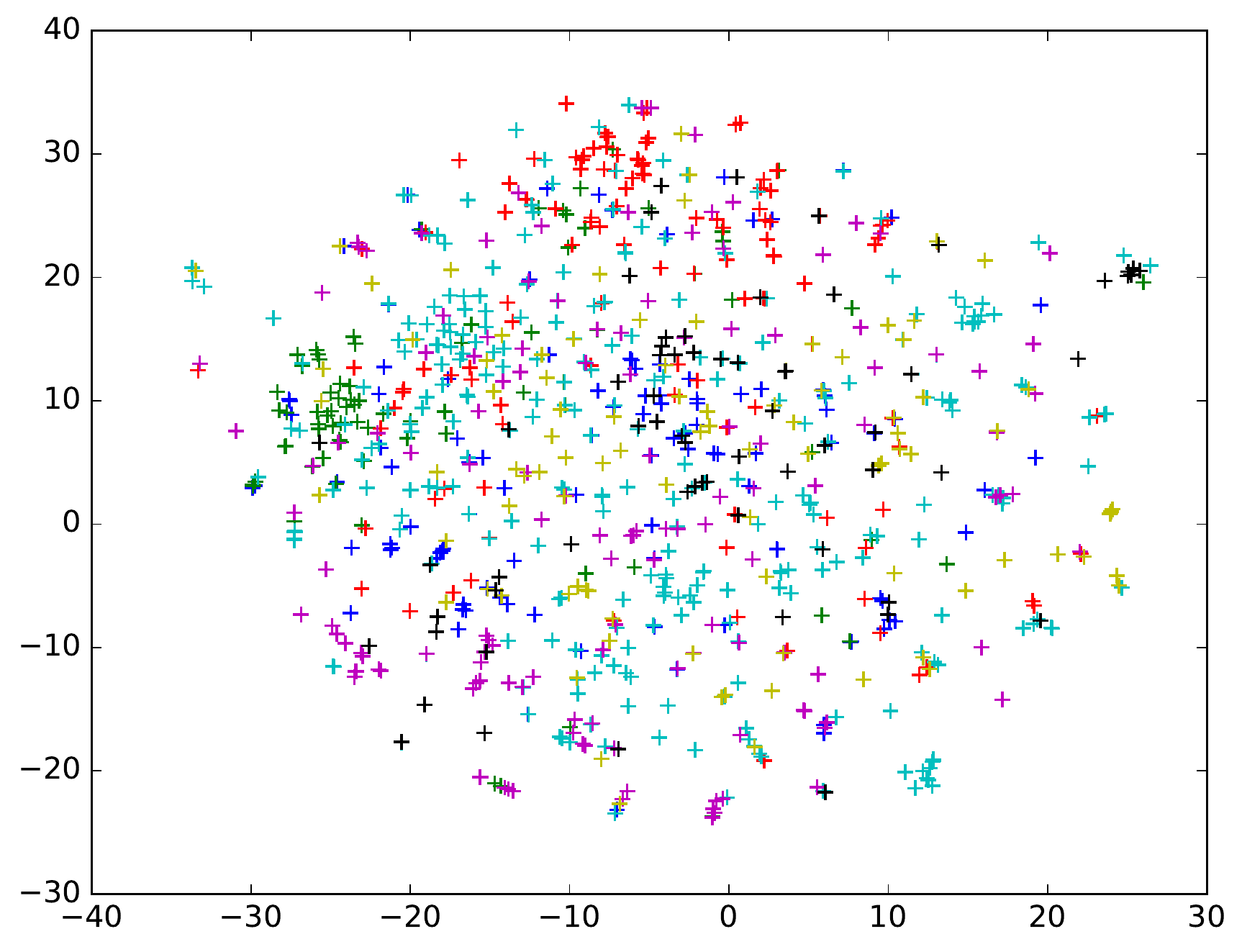}}
\vskip -0.1in
\caption{t-SNE visualization of embeddings on Cora}
\label{fig:visual}
\end{figure*}

We further visualize the embeddings learned by our unsupervised model \emph{APNE}
and two unsupervised embedding-based baselines
on the Cora dataset with a widely-used dimension reduction
method t-SNE~\cite{maaten2008visualizing}, and results are shown in Figure \ref{fig:visual}. 
One can observe that different classes are better separated by our model, and nodes in the same class are
clustered more tightly.

\begin{figure}[tb]
\centering
\centerline{\includegraphics[width=0.5\columnwidth]{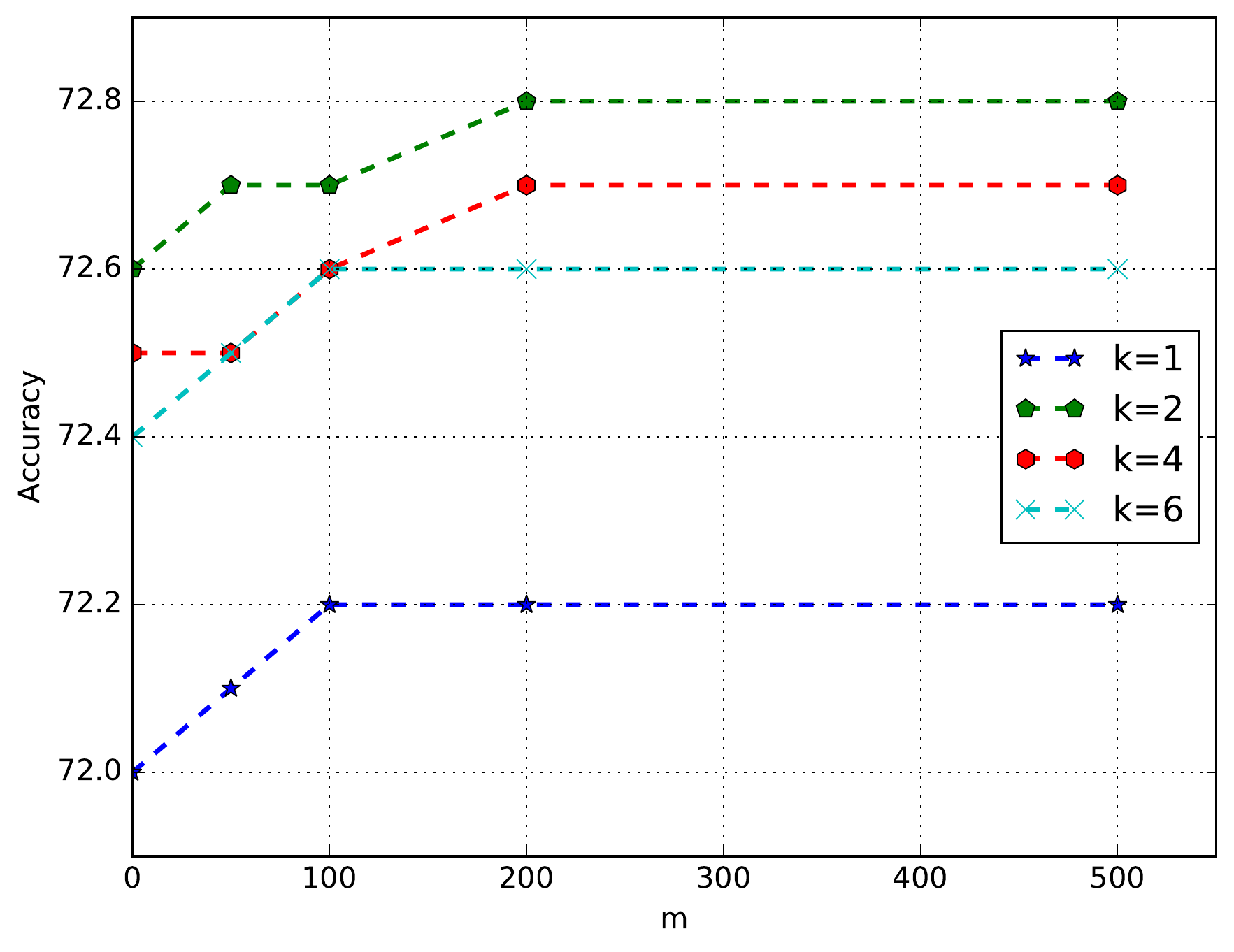}}
\vskip -0.1in
\caption{Parameter effect of \emph{APNE}  on the node classification task (in percentage)}
\label{fig:para}
\vskip -0.2in
\end{figure}

In order to test the sensitivity of our framework to hyper-parameters, we choose different values of
the negative sampling parameter $k$ in Theorem~\ref{theo:emf}
and the number of iterations of label context sampling $m$ in Algorithm~\ref{alg:A}
and evaluate the
model on Citeseer on the node classification task.

For each pair of parameters, we repeat the experiments 10 times and compute the mean accuracy.
Results are shown in Figure~\ref{fig:para}. The horizontal axis represents different values of $m$.
And $m=0$ represents
results when the model is purely unsupervised, otherwise results are from semi-supervised models.
The vertical axis is the classification accuracy on Citeseer. Clearly, increasing
$m$ brings a boost of the performance of the model, as we infer in Section \ref{sec:label}.
This justifies the effectiveness of the approach we propose to incorporate the label context.
In addition, the performance of the proposed models with different values of $k$ is relatively stable.

\subsection{Link Prediction}

\begin{table*}[ht]
\centering
\caption{Results of link prediction}
\label{tab:link}
\vskip -0.1in
\begin{tabular}{c||p{0.94cm} <{\centering}|p{0.94cm} <{\centering}||p{0.94cm} <{\centering}|p{0.94cm} <{\centering}||p{0.94cm} <{\centering}|p{0.94cm} <{\centering}||p{0.94cm} <{\centering}|p{0.94cm} <{\centering}}
\hline
{\multirow{2}{*}{Method}}
 & \multicolumn{2}{c||}{Citeseer} & \multicolumn{2}{c||}{Cora} & \multicolumn{2}{c||}{Pubmed} & \multicolumn{2}{c}{Facebook}\\
\cline{2-9}
\multirow{2}{*}{} & AUC & MAP & AUC & MAP & AUC & MAP & AUC & MAP \\
\hline
\hline
Common Neighbor & $0.567$ &$0.781$& $0.616$ &$0.797$& $0.561$ &$0.778$& $0.797$ &$0.882$\\
Jaccard's Coefficient & $0.567$&$0.782$ & $0.616$ &$0.795$& $0.561$ &$0.776$& $0.797$ &$0.877$\\
Adamic Adar & $0.560$ &$0.780$& $0.617$ &$0.801$& $0.561$ &$0.778$& $0.798$&$0.885$ \\
Preferential Attachment & $0.675$&$0.721$ & $0.679$ &$0.705$& $0.863$ &$0.852$& $0.675$&$0.675$ \\
\hline
DeepWalk & $0.656$&$0.725$ & $0.734$ &$0.793$& $0.721$ &$0.781$& $0.891$ &$0.914$\\
node2vec & $0.502$ &$0.731$& $0.723$ &$0.790$& $0.728$ &$0.785$& $0.888$ &$0.911$\\
TADW & $0.914$ &$0.936$& $0.854$ &$0.878$& $0.592$ &$0.620$& $0.909$ &$0.921$ \\
HSCA & $0.905$ & $0.928$ & $0.861$ & $0.885$ & $0.632$ & $0.660$ & $0.926$ & $0.917$ \\
\hline
\emph{APNE} & $\bm{0.938}$ &$\bm{0.940}$& $\bm{0.909}$ &$\bm{0.910}$& $\bm{0.925}$ &$\bm{0.916}$& $\bm{0.956}$ &$\bm{0.949}$\\
\hline
\end{tabular}
\end{table*}

We further test our unsupervised model on the link prediction task. In link prediction,
a snap-shot of the current network is given, and we are going to predict edges
that will be added in the future. 
The experiment is set up as follows: we first remove $50\%$ of existing edges
from the network randomly as positive node pairs,
while ensuring the residual network connected. To generate negative examples,
we randomly sample an equal number of node pairs that are not connected.
Node representations
are then learned based on the residual network. 
While testing, given a node pair in the samples,
we compute the cosine similarity between their representation vectors as the edge's score.
Finally, Area Under Curve (AUC) score and Mean Average Precision (MAP)
are used to evaluate the consistency between the labels
and the similarity scores of the samples.

Results are summarized in Table~\ref{tab:link}.
As shown in the table, our method \emph{APNE} outperforms all the baselines consistently
with different evaluation metrics.
We take a lead of topology-only methods by a large margin,
especially on sparser networks such as Citeseer, which indicates the
importance of leveraging node features on networks with high sparsity.
Again, we consistently outperform TADW and HSCA which also consider text features of nodes.

The stable performance of our proposed \emph{APNE} model on different datasets justify that
embeddings learned by jointly factorizing
the co-occurrence matrix $\bm{D}$ and node features $\bm{F}$
can effectively represent the network. More importantly, the problem of sparsity can be alleviated 
by incorporating node features in a unified framework.

\subsection{Case Study}

\begin{table}[tb]
\centering
\scriptsize
\caption{Two randomly chosen node pairs from Cora dataset}
\label{tab:case_mf}
\vskip -0.1in
\begin{tabular}{|p{6.5cm}|p{1.5cm} <{\centering}|p{1.5cm} <{\centering}|p{1cm} <{\centering}|p{1cm} <{\centering}|}
\hline
\multicolumn{1}{|c|}{\multirow{2}{*}{Title}} & \multirow{2}{*}{Same Class} & \multirow{2}{*}{Connected} & 
\multicolumn{2}{c|}{Cosine Similarity}\\
\cline{4-5}
\multirow{2}{*}{}&\multirow{2}{*}{}&\multirow{2}{*}{}&\emph{APNE}&TADW \\
\hline
A cooperative coevolutionary approach to function
& \multirow{2}{*}{$\surd$}
& \multirow{2}{*}{$\surd$}
& \multirow{2}{*}{$\bm{0.471}$}
& \multirow{2}{*}{$-0.002$}\\
\cline{1-1}
Multi-parent reproduction in genetic algorithms
& \multirow{2}{*}{}
& \multirow{2}{*}{}
& \multirow{2}{*}{}
& \multirow{2}{*}{}\\
\hline
A Class of Algorithms for Identification in $H_{\infty}$
& \multirow{2}{*}{$\surd$}
& \multirow{2}{*}{$\times$}
& \multirow{2}{*}{$\bm{0.158}$}
& \multirow{2}{*}{$-0.129$} \\
\cline{1-1}
On the Computational Power of Neural Nets
& \multirow{2}{*}{}
& \multirow{2}{*}{}
& \multirow{2}{*}{}
& \multirow{2}{*}{} \\
\hline
\end{tabular}
\vskip -0.1in
\end{table}

To further illustrate the effectiveness of \emph{APNE},
we present some instances of link prediction
on the Cora dataset. We randomly choose $2$ node pairs from all node samples
and compute the cosine similarity for each
pair. Results are summarized in Table~\ref{tab:case_mf}. 
The superiority of \emph{APNE} is obvious in the first instance,
where TADW gives a negative correlation
to a positive pair. For this pair, although the first paper is cited by the second one, their
neighbors do not coincide. 
As a consequence it is easy to wrongly separate these two nodes into different
categories
if the structure information is not sufficiently exploited.

As for the second instance, both papers belong to the Neural Networks class
but not connected in the network. 
Specifically,
the first paper focuses on H-Infinity methods in control theory while the second
paper is about recurrent neural networks, and there exist papers linking
these two
domains together in the dataset. 
As a consequence, although these two nodes can hardly co-occur in
random walk sequences on the network, their features may overlap in the dataset.
Therefore, the pair of nodes will have a higher feature similarity
than the topology similarity.
Thus by jointly considering
the network topology and the node features, our method gives a higher correlation
score to the two nodes that are disconnected but belong to the same category.

\section{Conclusion}
\label{sec:conclusion}
In this paper, we aim to learn a generalized network embedding preserving
structure, content and label information simultaneously. We propose a unified
matrix factorization based framework which provides a flexible
integration of network structure, node content, as well as label information.
We bridge the gap between word embedding and network embedding
by designing a method to generate the co-occurrence matrix from the network,
which is actually an approximation of high-order proximities of nodes in the network.
The experimental results on four benchmark datasets show that
the joint matrix factorization method we propose brings substantial
improvement over existing methods.
One of our future directions would be to apply our framework to
social recommendations to combine the relationship between users with the corresponding feature
representations.

\section*{Acknowledgements}
This research was supported by the National Natural Science Foundation of China (No. 61673364, No. U1605251 and No. 61727809), and the Fundamental Research Funds for the Central Universities (WK2150110008).

\bibliographystyle{splncs03}
\bibliography{dasfaa_98}

\end{document}